\documentclass[aps,prc,reprint,superscriptaddress,nofootinbib,floatfix]{revtex4-2}

\usepackage{amsmath}
\usepackage{amssymb}
\usepackage{bm}
\usepackage[dvipdfmx]{graphicx}
\usepackage{dcolumn}
\usepackage[dvipdfmx]{hyperref}
\usepackage[dvipdfmx]{xcolor}

\begin{document}

\title{Time-evolution formalism in the complex scaling method: Application to the E1 response of $^6$He}

\author{Yuma Kikuchi}
\email{kikuchi@tokuyama.ac.jp}
\affiliation{National Institute of Technology, Tokuyama College, Shunan 745-8585, Japan}
\affiliation{RIKEN Nishina Center, 2-1 Hirosawa, Wako 351-0198, Japan}

\author{Kiyoshi Kat\=o}
\affiliation{Nuclear Reaction Data Centre, Faculty of Science, Hokkaido University, Sapporo 060-0810, Japan}

\author{Takayuki Myo}
\affiliation{General Education, Faculty of Engineering, Osaka Institute of Technology 535-8585, Osaka, Japan}
\affiliation{Research Center for Nuclear Physics (RCNP), Osaka University, Ibaraki 567-0047, Japan}

\begin{abstract}
\begin{description}
\item[Background]
The complex scaling method (CSM) has been successfully used to describe many-body resonances as eigenvalues of the complex-scaled Hamiltonian in an appropriate $L^2$ basis representation. Its scope has subsequently been extended to many-body continuum states, strength functions, and scattering observables. However, a general framework that incorporates time evolution within the same CSM framework has not yet been established.

\item[Purpose]
We formulate a time-evolution formalism as a natural extension of the CSM based on the extended completeness relation (ECR), and apply it to the electric dipole (E1) excitation of $^6$He in order to clarify how an initially correlated three-body configuration evolves into continuum states.

\item[Methods]
Time evolution is described by a complex-scaled time-evolution operator represented with the ECR. The formalism is first tested in a simple two-body model through comparison with a direct numerical solution of the time-dependent Schr\"odinger equation. It is then applied to the E1 excitation of $^6$He in an $\alpha+n+n$ three-body model, and the density distributions are analyzed in different Jacobi coordinate systems.

\item[Results]
The present formalism reproduces the wave-packet evolution obtained in the direct time-dependent calculation. In the application to $^6$He, the initial E1-excited state exhibits a correlated configuration and evolves into spatially extended continuum states. The time evolution of the density distributions indicates the coexistence of sequential decay through a core-neutron subsystem and direct breakup.

\item[Conclusions]
The present formalism extends the scope of the CSM from spectral and scattering observables to real-time continuum dynamics, and provides a unified framework that connects initial-state correlations, continuum structure, and decay dynamics in weakly bound nuclei.
\end{description}
\end{abstract}

\keywords{non-Hermitian quantum mechanics, continuum states, few-body systems, halo nuclei, electric dipole transitions}

\maketitle

\section{Introduction}

Understanding the structure and dynamics of weakly bound nuclei is one of the central issues in nuclear physics. In such systems, coupling to the continuum plays an essential role, leading to exotic phenomena such as halo structures and characteristic excitation modes \cite{Tanihata1985,Hansen1987,Ikeda1992,Zhukov1993}. In particular, few-body descriptions are indispensable, because correlations among valence nucleons and their coupling to unbound states must be treated consistently.

The description of unbound states, including resonances and nonresonant continuum states, requires a theoretical framework beyond conventional bound-state approaches. Various methods based on non-Hermitian quantum mechanics and complex-energy representations have been developed \cite{Ho1983,Moiseyev1998}. These methods provide a natural framework for describing resonant states with finite lifetimes and their coupling to the continuum.

The complex scaling method (CSM) \cite{Aguilar1971,Balslev1971,Aoyama2006,Myo2014} provides a powerful framework for describing resonant states within an $L^2$ basis representation. By rotating the coordinates into the complex plane, resonant states with finite lifetimes appear as isolated eigenvalues of the complex-scaled Hamiltonian and can be clearly separated from the continuum spectrum. The method has been successfully applied to nuclear few-body systems and has been established as a powerful tool for the identification and description of resonances.

In addition to its original success in the description of resonances, the scope of the CSM has been extended to many-body continuum states and reaction observables. In particular, continuum responses can be formulated using complex-scaled Green's functions \cite{Myo2001,Myo2009}, and scattering observables can be described using the complex-scaled solutions of the Lippmann--Schwinger equation (CSLS) \cite{Kikuchi2009,Kikuchi2010}. These developments are particularly important for many-body systems, where continuum couplings, breakup channels, and scattering boundary conditions become highly nontrivial. Their success relies on the fact that the CSM provides a complete set of eigenstates, including many-body resonances and continuum states, through the extended completeness relation (ECR) \cite{Myo1997}. This property enables a consistent description of not only nuclear structure but also many-body continuum states, scattering states, and reaction observables within a unified framework.

Despite these advances, one important aspect has remained outside the established scope of the CSM: the explicit description of time evolution in the continuum. Time-dependent approaches have been successfully applied to decay dynamics in few-body systems, such as two-proton emission, where the time evolution of correlated particles provides important insights into decay mechanisms~\cite{Oishi2014}. However, a general formulation of such dynamical evolution within the CSM has not yet been fully established. Since the CSM has already been extended from resonance spectroscopy to continuum responses and scattering observables through the ECR, it is natural to ask whether the same spectral foundation can also be used to formulate quantum dynamics in the time domain. Establishing such a framework would broaden the scope of the CSM from a method formulated mainly in the energy representation to one that also provides a unified description of continuum dynamics.

In this work, we formulate a time-evolution formalism in the complex scaling method as a natural extension of the CSM, using the extended completeness relation (ECR) to construct a spectral representation of the complex-scaled time-evolution operator. The use of the ECR enables a unified treatment of bound, resonant, and nonresonant continuum states not only in the energy representation but also in the description of time evolution. We apply this formulation to the electric dipole (E1) excitation of $^6$He. The halo nucleus $^6$He is a typical three-body system consisting of an $\alpha$ core and two valence neutrons, in which continuum effects and spatial correlations play crucial roles in the excitation dynamics~\cite{Esbensen1997,Aumann1999}. By analyzing the time evolution of the E1-excited wave packet, we investigate how the initial correlated configuration evolves into continuum states and how different decay modes emerge in the three-body system. In this way, the present formalism provides a direct connection among nuclear structure, continuum structure, and decay dynamics within a unified CSM framework.

The paper is organized as follows. In Sec.~II, we present the time-evolution formalism in the complex scaling method. In Sec.~III, we describe its application to a test calculation and to the E1 excitation of $^6$He within a three-body model. In Sec.~IV, we present the results and discuss the time evolution of the density distributions in different Jacobi coordinate systems. Finally, Sec.~V gives a summary.

\section{Formulation}

\subsection{Complex scaling method}

In the complex scaling method (CSM)~\cite{Aguilar1971,Balslev1971}, the spatial coordinates of a quantum system are rotated into the complex plane according to
\begin{equation}
\mathbf{r} \rightarrow \mathbf{r} e^{i\theta},
\end{equation}
where $\theta$ denotes the real scaling angle. This transformation is implemented through a similarity transformation generated by the operator $\hat{U}(\theta)$, which performs the complex rotation of the coordinates. Applying this transformation to the Hamiltonian operator $\hat{H}$, the complex-scaled Hamiltonian is defined as
\begin{equation}
\hat{H}^{\theta}
=
\hat{U}(\theta)\,\hat{H}\,\hat{U}^{-1}(\theta),
\end{equation}
while the original Hamiltonian can be written as
\begin{equation}
\hat{H}
=
\hat{U}^{-1}(\theta)\,\hat{H}^{\theta}\,\hat{U}(\theta).
\end{equation}

To obtain the eigenstates of $\hat H^\theta$, we solve the eigenvalue problem using a set of square-integrable ($L^2$) basis functions~\cite{Myo2014}. In the present work, the eigenstates are expanded in this basis and obtained through diagonalization. As a result, in addition to bound states, components corresponding to resonant and continuum states are obtained as discretized eigenstates. The eigenvalue problem is written as
\begin{equation}
\hat H^\theta |\Psi_\nu^\theta\rangle = E_\nu^\theta |\Psi_\nu^\theta\rangle .
\end{equation}
Here, $|\Psi_\nu^\theta\rangle$ denotes a discretized eigenstate with state index $\nu$ obtained within the $L^2$ basis representation, and $E_\nu^\theta$ is the corresponding complex eigenvalue.

Since $\hat{H}^\theta$ is generally non-Hermitian, the corresponding left eigenstates $\langle \widetilde{\Psi}_\nu^\theta |$ are introduced to construct a biorthogonal set. These states satisfy the biorthogonality relation
\begin{equation}
\langle \widetilde{\Psi}_\nu^\theta | \Psi_\mu^\theta \rangle = \delta_{\nu\mu} .
\label{eq:biorthogonality}
\end{equation}
Using this biorthogonal basis, the identity operator is represented by the extended completeness relation (ECR),
\begin{equation}
\sum_\nu |\Psi_\nu^\theta\rangle \langle \widetilde{\Psi}_\nu^\theta| = \mathbf{1}^\theta ,
\label{eq:ECR}
\end{equation}
where $\mathbf{1}^\theta$ denotes the identity operator in the function space spanned by the eigenstates of $\hat{H}^\theta$ within the $L^2$ basis representation. Although the continuum spectrum is discretized in practical calculations, the ECR provides a useful spectral representation in which bound, resonant, and nonresonant continuum states are treated on the same footing. Its validity and usefulness have been demonstrated in applications such as the calculation of the continuum level density~\cite{Suzuki2005} and strength functions based on the complex-scaled Green's function~\cite{Myo2001}, where physical observables become essentially independent of $\theta$ for sufficiently large scaling angles. This property forms the basis of the present formulation of time evolution within the CSM framework.

\subsection{Spectral representation of the complex-scaled time-evolution operator}

Using the same extended completeness relation (ECR) that has been employed in CSM descriptions of continuum responses and scattering states, we construct a spectral representation of the time-evolution operator. In this sense, the present formalism should be viewed as a natural extension of the CSM to the explicit description of quantum dynamics in the continuum.

The time evolution of a quantum state is governed by the time-evolution operator generated by the Hamiltonian $\hat{H}$,
\begin{equation}
|\Psi(t)\rangle
=
e^{-\frac{i}{\hbar}\hat{H}t}
|\Psi(0)\rangle.
\end{equation}
Using the relation between $\hat{H}$ and $\hat{H}^{\theta}$, this operator can be written as
\begin{equation}
e^{-\frac{i}{\hbar}\hat{H}t}
=
\hat{U}^{-1}(\theta)\,
e^{-\frac{i}{\hbar}\hat{H}^{\theta}t}\,
\hat{U}(\theta).
\end{equation}

By inserting the ECR, the complex-scaled time-evolution operator is expanded as
\begin{equation}
e^{-\frac{i}{\hbar}\hat{H}^{\theta}t}
=
\sum_{\nu}
|\Psi_\nu^{\theta}\rangle
\, e^{-\frac{i}{\hbar}E_\nu^{\theta}t} \,
\langle \widetilde{\Psi}_\nu^{\theta} |,
\end{equation}
and the full time-evolution operator becomes
\begin{equation}
e^{-\frac{i}{\hbar}\hat{H}t}
=
\hat{U}^{-1}(\theta)
\left(
\sum_{\nu}
|\Psi_\nu^{\theta}\rangle
\, e^{-\frac{i}{\hbar}E_\nu^{\theta}t} \,
\langle \widetilde{\Psi}_\nu^{\theta} |
\right)
\hat{U}(\theta).
\end{equation}

For an initial state $|\Phi(0)\rangle$, the wave function at time $t$ is given by
\begin{equation}
|\Phi(t)\rangle
=
e^{-\frac{i}{\hbar}\hat{H}t}
|\Phi(0)\rangle,
\end{equation}
which leads to
\begin{equation}
|\Phi(t)\rangle
=
\hat{U}^{-1}(\theta)
\sum_{\nu}
|\Psi_\nu^{\theta}\rangle
\, e^{-\frac{i}{\hbar}E_\nu^{\theta}t} \,
\langle \widetilde{\Psi}_\nu^{\theta} |
\hat{U}(\theta)
|\Phi(0)\rangle.
\label{eq:te_wf}
\end{equation}

For resonance states, the complex eigenvalue is written as
\begin{equation}
E_\nu^{\theta}
=
E_r - i\Gamma/2,
\end{equation}
where $E_r$ and $\Gamma$ denote the resonance energy and decay width, respectively. The corresponding time dependence becomes
\begin{equation}
e^{-\frac{i}{\hbar}E_r t}
\, e^{-\frac{\Gamma}{2\hbar} t},
\end{equation}
which naturally describes the exponential decay of resonance states.

For nonresonant continuum states, the complex-scaled energy eigenvalues are distributed along the $2\theta$-rotated lines in the lower half complex-energy plane. As a result, each discretized continuum component acquires an exponential damping factor in its time evolution. This damping factor, however, does not represent a physical decay width, but rather originates from the complex scaling.
This feature can be understood as being equivalent to an absorbing boundary condition for the outgoing flux. In this sense, the CSM plays a role analogous to that of absorbing boundary conditions or complex absorbing potentials, which are commonly introduced in time-dependent scattering calculations to suppress reflections of the outgoing flux~\cite{Ueda2003}.

\section{Applications}

\subsection{Test calculation}

To demonstrate the validity of the present time-evolution formalism in the complex scaling method, we first perform a test calculation for a simple model system and compare the results with a direct numerical solution of the time-dependent Schr\"odinger equation obtained by the Crank--Nicolson method.

For simplicity, we consider only the radial part of the Hamiltonian and the wave function. The Hamiltonian for the radial motion with orbital angular momentum $l$ is written as
\begin{equation}
\hat H=
-\frac{\hbar^2}{2\mu}
\left(
\frac{d^2}{dr^2}
-\frac{l(l+1)}{r^2}
\right)
+V(r).
\label{eq:test_hamiltonian}
\end{equation}
For simplicity, the reduced mass $\mu$ is set equal to the nucleon mass. The interaction is taken to be a one-range Gaussian potential,
\begin{equation}
V(r)=V_0 e^{-\beta r^2},
\label{eq:test_potential}
\end{equation}
with the parameters
\begin{equation}
V_0=-25~\mathrm{MeV},
\qquad
\beta=\frac{1}{9}~\mathrm{fm}^{-2}.
\label{eq:test_parameters}
\end{equation}
We consider the time evolution of a wave packet in the $l=1$ ($p$-wave) case. In this model, the $p$-wave interaction generates a resonance at $E_r=0.377$ MeV with a decay width $\Gamma=0.342$ MeV, indicating the presence of a relatively broad resonance.

As the initial state, we employ a Gaussian wave packet with $p$-wave radial dependence. The radial part of the initial wave function is taken as
\begin{equation}
u(r,0)=N r e^{-r^2/8},
\label{eq:test_initial}
\end{equation}
where $N$ is a normalization constant. The corresponding state vector is denoted by $|\Phi(0)\rangle$, and its radial coordinate representation at time $t$ is written as
\begin{equation}
u(r,t)=\langle r|\Phi(t)\rangle.
\label{eq:test_coordinate_representation}
\end{equation}

The time evolution of the wave packet is evaluated using the general expression in Eq.~\eqref{eq:te_wf}.
In the present test calculation, the eigenstates $\{\Psi_\nu^\theta\}$ are obtained by diagonalizing the complex-scaled Hamiltonian in a Gaussian basis. Since we consider the $l=1$ partial wave, the radial wave functions are expanded in Gaussian basis functions with a $p$-wave radial dependence,
\begin{equation}
\phi_n(r)=N_n\, r \exp\left(-\frac{r^2}{2b_n^2}\right),
\label{eq:test_basis}
\end{equation}
where $N_n$ is the normalization constant and $b_n$ is the range parameter. In the present calculation, 60 Gaussian basis functions are employed, and the range parameters $b_n$ are chosen in geometric progression from 0.01 fm to 60 fm. This choice provides an efficient description of both the short-range structure and the long-range tail of the wave packet within a finite basis space. The complex scaling angle is taken to be $10$ degrees. This procedure provides a discrete set of eigenstates, including both resonance and continuum components, which are used in the spectral representation of the time evolution.

To examine the time evolution in a physically transparent way, we introduce the radial density distribution
\begin{equation}
\rho(r,t)=|u(r,t)|^2 .
\label{eq:test_density}
\end{equation}
Here, \(u(r,t)\) is the reduced radial wave function, and \(\rho(r,t)\) therefore represents the probability distribution in the radial coordinate. This quantity provides a simple measure of how the wave packet propagates outward and spreads in coordinate space. In addition, we consider the integrated norm
\begin{equation}
N(t)=\int_0^\infty dr\,\rho(r,t),
\label{eq:test_norm}
\end{equation}
whose time dependence characterizes the absorptive nature of the complex-scaled representation.

By substituting the general expression in Eq.~\eqref{eq:te_wf} into the coordinate-space representation of the wave function, we obtain
\begin{equation}
u(r,t)=\sum_\nu \varphi^\theta_\nu(r)e^{-iE_\nu^\theta t/\hbar}c^\theta_\nu ,
\label{eq:test_u_spectral}
\end{equation}
where
\begin{equation}
\varphi^\theta_\nu(r)=\langle r|U^{-1}(\theta)|\Psi_\nu^\theta\rangle,
\qquad
c^\theta_\nu=\langle \widetilde{\Psi}_\nu^\theta|U(\theta)|\Phi(0)\rangle .
\label{eq:test_coefficients}
\end{equation}
Here, \(\varphi^\theta_\nu(r)\) denotes the back-rotated coordinate-space component of the complex-scaled eigenstate, while \(c^\theta_\nu\) is the expansion coefficient of the initial state in the biorthogonal basis of the complex-scaled Hamiltonian. In this representation, the physical wave function is written as a superposition of resonant and discretized nonresonant continuum components.

Accordingly, the density distribution in Eq.~\eqref{eq:test_density} is written as
\begin{equation}
\rho(r,t)=\sum_{\nu\nu'}
\varphi^\theta_\nu(r)\varphi_{\nu'}^{\theta *}(r)
e^{-i(E_\nu^\theta-E_{\nu'}^{\theta *})t/\hbar}
c^\theta_\nu c_{\nu'}^{\theta *}.
\end{equation}
This expression shows that the density contains not only diagonal contributions from individual eigenstates but also interference terms between different components. The time evolution of \(\rho(r,t)\) therefore reflects the interplay between resonant and nonresonant continuum contributions in the propagation of the wave packet.

\begin{figure}[t]
\centering
\includegraphics[width=0.9\linewidth]{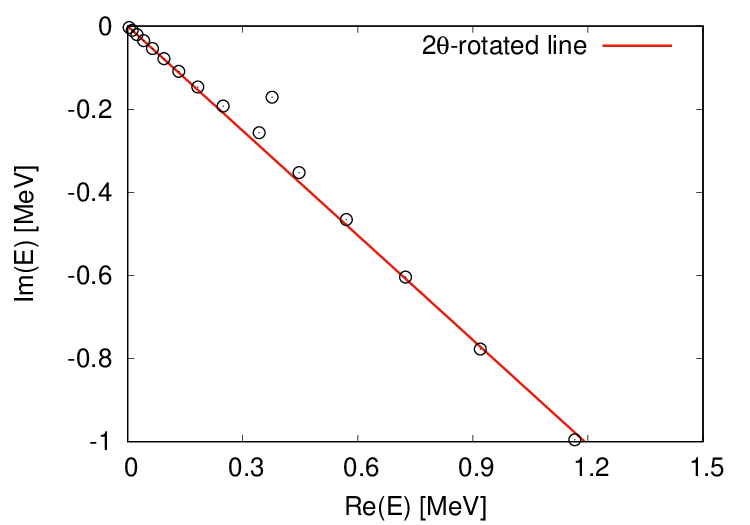}
\caption{
Distribution of the complex energy eigenvalues for the model Hamiltonian, obtained with 60 Gaussian basis functions whose range parameters are taken in geometric progression from 0.01 fm to 60 fm. The scaling angle is taken to be $\theta=20$ degrees. The open circles represent the discretized eigenvalues obtained with the $L^2$ basis, and the solid line indicates the $2\theta$-rotated branch cut. The isolated pole corresponds to the resonance at $E_r-i\Gamma/2 = 0.377 - 0.172\,i$ MeV.
}
\label{fig:edist_2body}
\end{figure}

Figure~\ref{fig:edist_2body} shows the distribution of the complex energy eigenvalues obtained for the model Hamiltonian. The isolated pole corresponds to the resonant state, whereas the other eigenvalues are aligned along the $2\theta$-rotated branch cut and represent the discretized nonresonant continuum states. This eigenvalue distribution is consistent with the general spectral structure of the CSM and provides the basis for the spectral representation of the time evolution used in the present work.

For comparison, the time-dependent Schr\"odinger equation
\begin{equation}
i\hbar \frac{\partial}{\partial t}u(r,t)=\hat{H}u(r,t)
\label{eq:test_tdse}
\end{equation}
is also solved directly in coordinate space using the Crank--Nicolson method. By comparing the time evolution of \(\rho(r,t)\) obtained from the present formalism with that obtained from the direct solution of Eq.~(27), we examine the validity of the present time-evolution formalism in the CSM.

In the present formalism, \(N(t)\) is expected to decrease with time because complex scaling effectively introduces an absorbing boundary condition for outgoing components. This reduction should not be interpreted as a physical loss of probability. Rather, it reflects the damping of outgoing flux inherent in the complex-scaled representation. In this sense, the time dependence of \(N(t)\) provides a measure of the effective absorption associated with the complex scaling transformation. We also examine the dependence of \(N(t)\) on the scaling angle \(\theta\) to confirm that this behavior is essentially stable against the choice of \(\theta\).

\subsection{Application to the E1 excitation of $^{6}$He}

\subsubsection{Three-body model of $^{6}$He}

The halo nucleus $^{6}$He is described using a three-body model consisting of an $\alpha$ core and two valence neutrons. In the present work, the system is treated within the framework of the orthogonality condition model (OCM)~\cite{Saito1969}, which effectively incorporates the Pauli principle between the valence neutrons and the nucleons inside the $\alpha$ core.

The Hamiltonian of the system is written as
\begin{equation}
\hat{H}
=
\hat{T}
+
\hat{V}_{n\alpha}^{(1)}
+
\hat{V}_{n\alpha}^{(2)}
+
\hat{V}_{nn}
+
\hat{V}_{3b}
+
\hat{V}_{\mathrm{Pauli}} .
\end{equation}
Here $\hat{T}$ denotes the kinetic energy operator for the relative motion of the three-body system.
The operators $\hat{V}^{(1)}_{n\alpha}$ and $\hat{V}^{(2)}_{n\alpha}$ represent the interactions between each
valence neutron and the $\alpha$ core, while $\hat{V}_{nn}$ denotes the neutron--neutron interaction.
For the neutron--$\alpha$ interaction we employ the KKNN potential~\cite{Kanada1979}, which reproduces the low-energy $n$--$\alpha$ scattering properties.
For the neutron--neutron interaction we use the Minnesota force~\cite{Thompson1977}, which is used as an effective nucleon--nucleon interaction in few-body calculations.

In addition to these two-body interactions, we introduce a phenomenological three-body cluster interaction
$\hat{V}_{3b}$.
The functional form and parameters of this interaction are taken from Ref.~\cite{Kikuchi2010}, where it has been shown to reproduce the ground-state properties of $^6$He such as the binding energy and matter radius.

Within the OCM framework, the Pauli forbidden states between the valence neutrons and the $\alpha$ core are eliminated by introducing a pseudo potential~\cite{Krasnopolsky1974,Kukulin1976}
\begin{equation}
\hat{V}_{\mathrm{Pauli}}
=
\lambda
\sum_{f}
|u_{f}\rangle
\langle u_{f}| ,
\end{equation}
where $|u_{f}\rangle$ denotes the Pauli-forbidden states and $\lambda$ is taken to be sufficiently large.
In the present calculation, we assume that the $\alpha$ core is inert, and thus, the $u_f$ is the $0s$ orbit for the $\alpha$-$n$ relative motion.

The eigenstates of the three-body Hamiltonian are obtained by solving the complex-scaled Schr\"odinger equation
\begin{equation}
\hat{H}^\theta|\Psi^\theta_\nu\rangle
=
E^\theta_\nu |\Psi^\theta_\nu\rangle .
\label{eq:csm_schrodinger}
\end{equation}
In the present calculation, Eq.~(\ref{eq:csm_schrodinger}) is solved using the Gaussian expansion method (GEM)~\cite{Hiyama2003}. The three-body wave function with total angular momentum $J$ and its projection $M$ is expanded in each Jacobi coordinate set as
\begin{equation}
\Psi_{JM}
=
\sum_{i}
C_{i}
\,
\mathcal{A}
\Bigl[
\bigl[
\phi_{l_1}(a_{n_1};\mathbf r_c)
\otimes
\phi_{l_2}(a_{n_2};\mathbf R_c)
\bigr]_{L}
\otimes
\chi_{S}
\Bigr]_{JM},
\end{equation}
where $i=\{c,n_1,n_2,l_1,l_2,L,S\}$, $l_1$ and $l_2$ denote the orbital angular momenta associated with the Jacobi coordinates $\mathbf r_c$ and $\mathbf R_c$, respectively, and $\chi_S$ denotes the spin wave function of the two valence neutrons coupled to total spin $S$. The antisymmetrizer between the two valence neutrons is denoted by $\mathcal{A}$. The Gaussian basis function is defined by
\begin{equation}
\phi_l(a_n;\mathbf r)
=
N_l(a_n)\,
r^l
\exp\!\left(-\frac{1}{2}a_n r^2\right)
Y_l(\hat{\mathbf r}),
\end{equation}
with $a_n = 1/b_n^2$ and $b_n=b_0\gamma^{n-1}$. The same Gaussian parameter set $\{a_n\}$ is used for both relative coordinates, and the range parameters $b_n$ are chosen in geometric progression from $0.3$ fm to $40$ fm. In the present model space, 25 Gaussian basis functions are employed for each coordinate, with partial waves up to $l_1,l_2 \le 3$. The scaling angle is taken to be $\theta = 10$~degrees.
The expansion coefficients $\{C_i\}$ are determined by diagonalizing the complex-scaled Hamiltonian matrix.
The numerical convergence of the results has been confirmed with respect to the basis size and the model space.

\begin{figure}[t]
\centering
\includegraphics[width=0.9\linewidth]{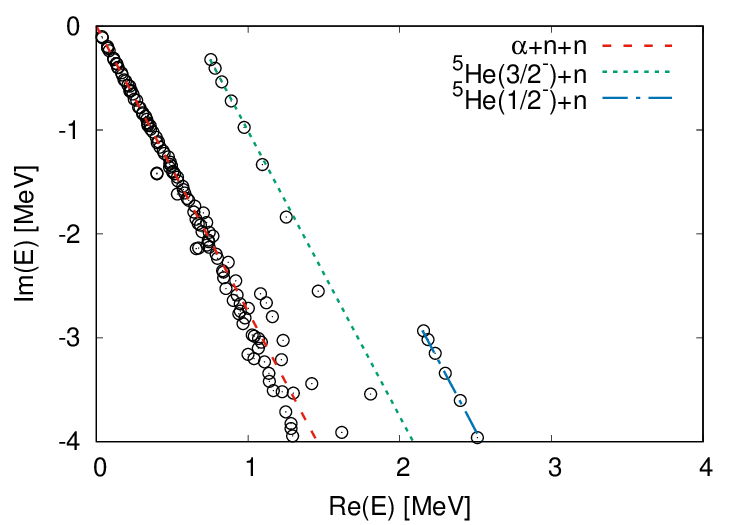}
\caption{Distribution of the complex energy eigenvalues for the $1^-$ states of $^{6}$He obtained with the scaling angle $\theta = 35$ degrees. The open circles represent the discretized eigenvalues obtained with the $L^2$ basis. The dashed, dotted, and dash-dotted lines indicate the $2\theta$-rotated branch cuts corresponding to the $\alpha+n+n$, $^{5}$He$(3/2^-)+n$, and $^{5}$He$(1/2^-)+n$ channels, respectively. No isolated resonance poles are found for the present $1^-$ calculation.}
\label{fig:edist}
\end{figure}
The eigenvalue distribution shown in Fig.~\ref{fig:edist} is presented for illustrative purposes to clarify the spectral structure of the complex-scaled Hamiltonian. The model space and parameters used in this figure are slightly simplified compared to those employed in the full three-body calculations described above. Nevertheless, the essential features of the continuum spectrum, such as the rotated branch cuts and the absence of isolated resonance poles, are preserved.

\subsubsection{E1 excitation and time evolution}

The electric dipole operator induces transitions from the ground state to the continuum states of the system. For the three-body $\alpha+n+n$ system, the E1 operator is written as
\begin{equation}
\hat{O}_{E1}
=
e\,\frac{2Z_c}{A}\,R_3 Y_1(\hat{\mathbf{R}}_3).
\end{equation}
Here $Z_c$ is the atomic number of the core, $A$ is the total mass number of the system, and $\mathbf{R}_3$ denotes the Jacobi coordinate corresponding to the relative motion between the $\alpha$ core and the center of mass of the two neutrons.

The initial state immediately after the dipole excitation is defined as
\begin{equation}
|\Phi(0)\rangle
=
\hat{O}_{E1}
|\Psi_0\rangle ,
\label{eq:init_6He}
\end{equation}
where $|\Psi_0\rangle$ denotes the ground-state wave function of $^{6}$He.
The subsequent time evolution of the E1-excited state is described using the general expression in Eq.~\eqref{eq:te_wf}, with the initial state given by Eq.~\eqref{eq:init_6He}.

\begin{figure}[t]
\centering
\includegraphics[width=0.9\linewidth]{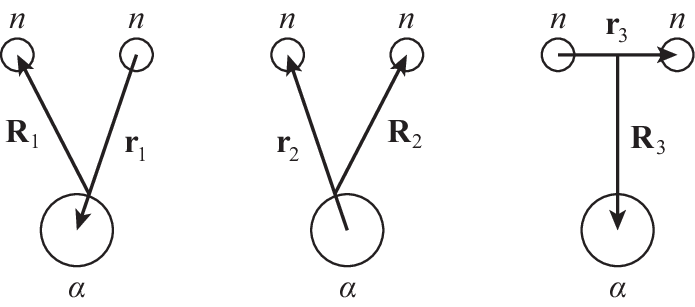}
\caption{Schematic illustration of the Jacobi coordinate systems used in the present work.}
\label{fig:Jacobi}
\end{figure}

To investigate the spatial structure of the wave packet, the wave function is expressed in the coordinate representation of the Jacobi coordinates
\begin{equation}
\Phi(\mathbf r_i,\mathbf R_i,t)
=
\langle
\mathbf r_i,\mathbf R_i
|
\Phi(t)
\rangle ,
\end{equation}
where $\mathbf r_i$ and $\mathbf R_i$ denote the Jacobi coordinates of the three-body system.
A schematic illustration of the Jacobi coordinate system used in the present work is shown in Fig.~\ref{fig:Jacobi}.

\subsubsection{Time evolution of the density distribution}

The spatial structure of the evolving wave packet is investigated using the density distribution
\begin{equation}
\rho(\mathbf r_i,\mathbf R_i,t)
=
\left|
\Phi(\mathbf r_i,\mathbf R_i,t)
\right|^{2}.
\end{equation}

To visualize the spatial configuration of the three-body system, the angular variables are integrated out, and a two-dimensional density distribution is obtained in terms of the radial coordinates
\begin{equation}
r_i = |\mathbf r_i|,
\qquad
R_i = |\mathbf R_i|.
\end{equation}
The resulting density distribution is defined as
\begin{equation}
D(r_i,R_i,t)
=
r_i^{2} R_i^{2}
\int
d\Omega_{r_i}\,
d\Omega_{R_i}
\left|
\Phi(\mathbf r_i,\mathbf R_i,t)
\right|^{2}.
\end{equation}
This quantity allows us to visualize the time evolution of the spatial correlations in the E1-excited three-body continuum.

\section{Results and discussion}

\subsection{Test calculation}

\begin{figure}[t]
\centering
\includegraphics[width=0.8\linewidth]{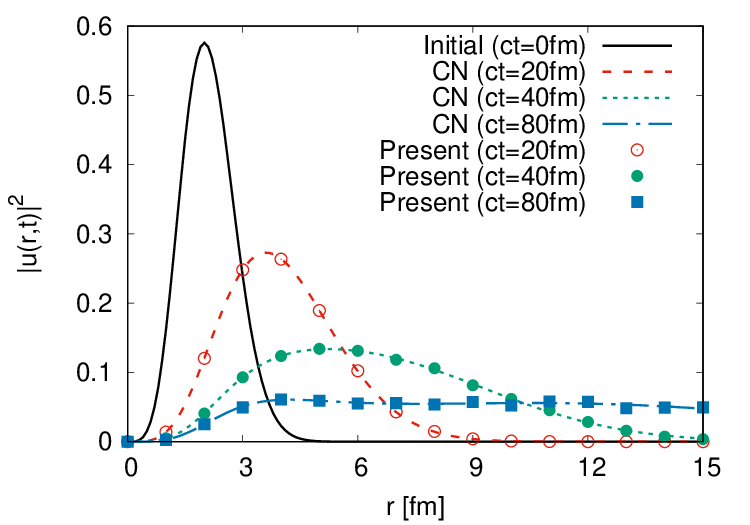}
\caption{
Time evolution of the radial density distribution $\rho(r,t)=|u(r,t)|^2$ for the test calculation. The solid black line indicates the initial wave packet at $ct = 0~\mathrm{fm}$. The results obtained using the present time-evolution formalism are shown by open red circles, closed green circles, and blue squares at $ct = 20~\mathrm{fm}$, $40~\mathrm{fm}$, and $80~\mathrm{fm}$, respectively. These results are compared with those obtained by the Crank--Nicolson (CN) method, shown by the dashed red, dotted green, and dash-dotted blue lines at the corresponding times.
The density distribution is given in units of fm$^{-1}$.
}
\label{fig:test}
\end{figure}

We first present the results of the test calculation in order to validate the present time-evolution formalism in the complex scaling method. The time evolution of the Gaussian wave packet obtained from the present spectral expansion is compared with the direct numerical solution of the time-dependent Schr\"odinger equation obtained by the Crank--Nicolson method. The comparison is performed for the radial density distribution $\rho(r,t)=|u(r,t)|^{2}$,
and the results are shown in Fig.~\ref{fig:test}.

In the present model, the $p$-wave interaction generates a resonance state with a resonance energy of 0.377 MeV and a decay width of 0.342 MeV, indicating that the system contains a relatively broad resonant state. The time evolution is examined at several representative times, $ct = 20$, $40$, and $80~\mathrm{fm}$. At all of these times, the wave packet obtained by the present formulation agrees well with that obtained by the Crank--Nicolson method, successfully reproducing both the outward propagation and the gradual spreading of the density distribution.

At the later time $ct = 80~\mathrm{fm}$, a component associated with the resonant state remains visible in the inner region, while the nonresonant continuum components propagate toward larger distances. The oveall agreement between the two methods demonstrates that the present formulation provides an accurate description of the time evolution of continuum wave packets, including contributions from a broad resonance.

\begin{figure}[t]
\centering
\includegraphics[width=0.8\linewidth]{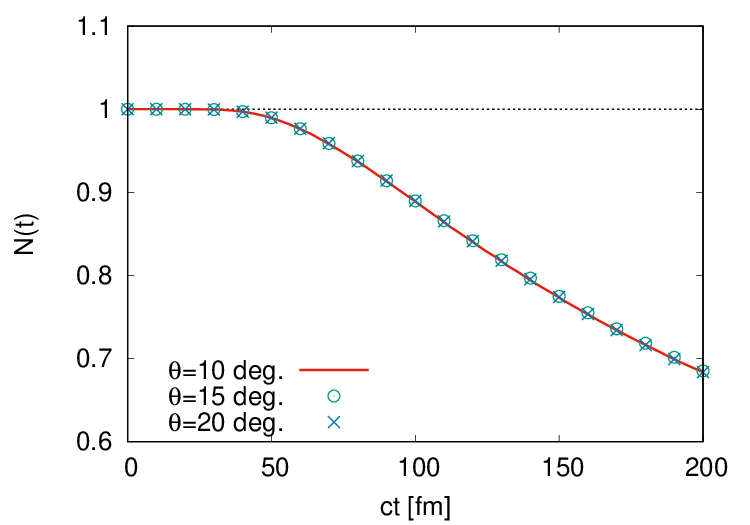}
\caption{
Time dependence of the integrated norm $N(t)$ in the test calculation.
The solid red line, open green circles, and  blue crosses show the results for $\theta=10$, $15$, and $20$ degrees, respectively,
and the dotted black line indicates $N(t)=1$.
}
\label{fig:norm}
\end{figure}
We also examine the time dependence of the integrated norm $N(t)$ as shown in Fig.~\ref{fig:norm}.
The norm decreases with time, reflecting the absorbing character of the complex scaling transformation. As discussed in Sec.~II, the discretized nonresonant continuum eigenvalues are distributed along the rotated branch cut in the lower half of the complex-energy plane, and each component acquires an exponential damping factor in time evolution. Accordingly, the decrease of $N(t)$ should be understood not as a physical loss of probability but as a manifestation of the effective absorption of outgoing flux in the complex-scaled representation. Furthermore, we confirm that the time dependence of $N(t)$ is almost independent of the scaling angle $\theta$, indicating that the damping behavior is not an artifact of a particular choice of $\theta$ but a robust feature of the present formulation.

\subsection{Time evolution of the density distribution}

Using the wave function defined in Sec.~III-B, we evaluate the time evolution of the density distribution $D(r_i,R_i,t)$ of $^6$He.
In the following, we denote the set $(r_3,R_3)$ as $(r_{n\text{-}n},R_{\alpha\text{-}nn})$, where $r_{n\text{-}n}$ represents the relative distance between the two neutrons and $R_{\alpha\text{-}nn}$ denotes the distance between the $\alpha$ core and the center of mass of the two neutrons.
Similarly, the set $(r_1,R_1)$ is denoted as $(r_{\alpha\text{-}n},R_{n\text{-}\alpha n})$, where $r_{\alpha\text{-}n}$ represents the relative distance between the $\alpha$ core and one neutron, and $R_{n\text{-}\alpha n}$ denotes the distance between the remaining neutron and the center of mass of the core-neutron subsystem.

\begin{figure}[t]
\centering
\includegraphics[trim=30 0 30 0, width=0.7\linewidth]{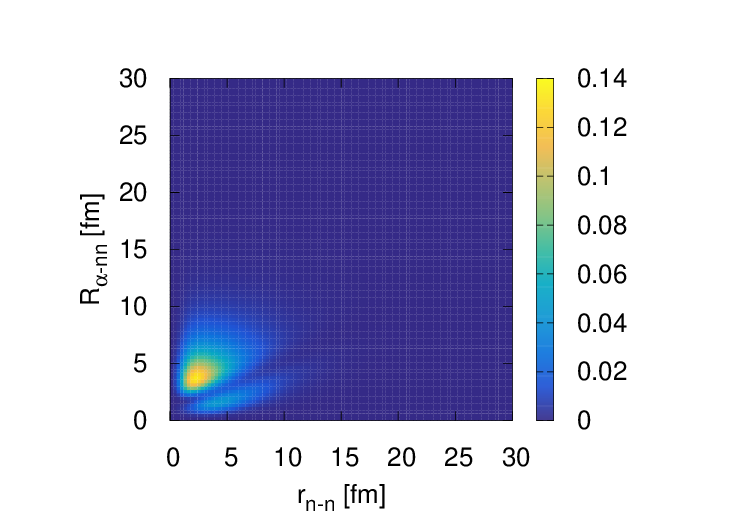}

\vspace{0.3cm}

\includegraphics[trim=30 0 30 0, width=0.7\linewidth]{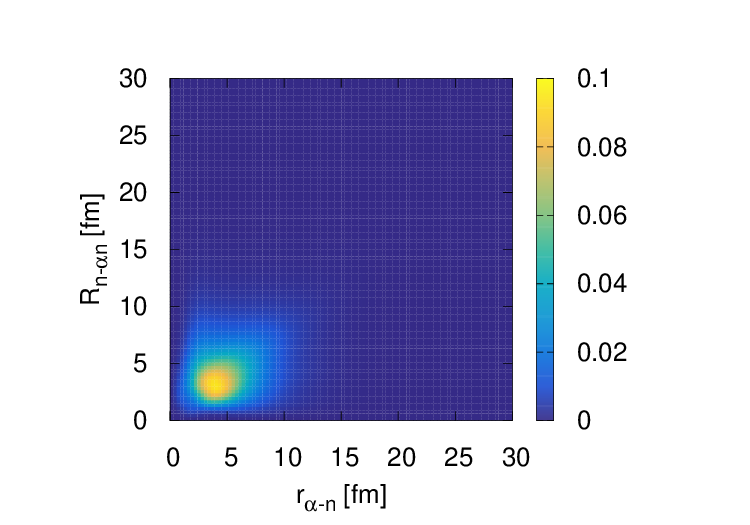}
\caption{
Density distributions of the initial state of the E1-excited $^{6}\mathrm{He}$ system in two different Jacobi coordinate systems (units in $e^2$). (Upper) Density distribution in the coordinate set describing the relative motion of the two neutrons ($r_{n\text{-}n}, R_{\alpha\text{-}nn}$). (Lower) Density distribution in the coordinate set corresponding to the core--neutron relative motion ($r_{\alpha\text{-}n}, R_{n\text{-}\alpha n}$).
}
\label{fig:init_density}
\end{figure}

We first examine the density distributions of the initial state of the E1-excited $^6$He in two different Jacobi coordinate systems, as shown in Fig.~\ref{fig:init_density}. In the coordinate set describing the relative motion of the two neutrons, the density distribution is concentrated in the region of small $r_{n\text{-}n}$, indicating that the initial E1 excitation predominantly populates dineutron-like configurations. In contrast, in the coordinate set corresponding to the core--neutron relative motion, the density distribution exhibits a single peak structure, reflecting the spatial correlation of the three-body system in the initial state.

\begin{figure*}[t]
\centering
\includegraphics[trim=30 0 30 0, width=0.32\textwidth]{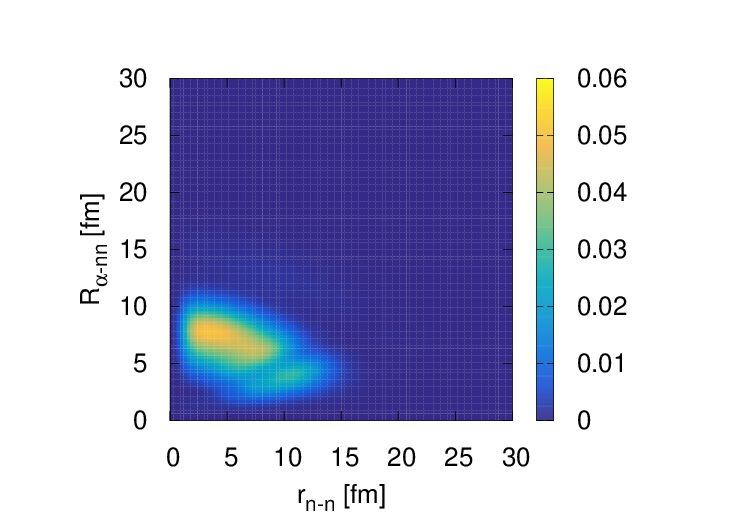}
\includegraphics[trim=30 0 30 0, width=0.32\textwidth]{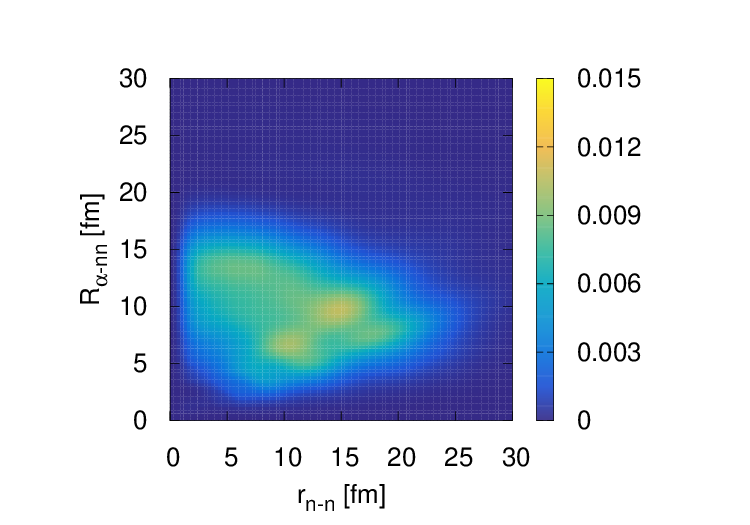}
\includegraphics[trim=30 0 30 0, width=0.32\textwidth]{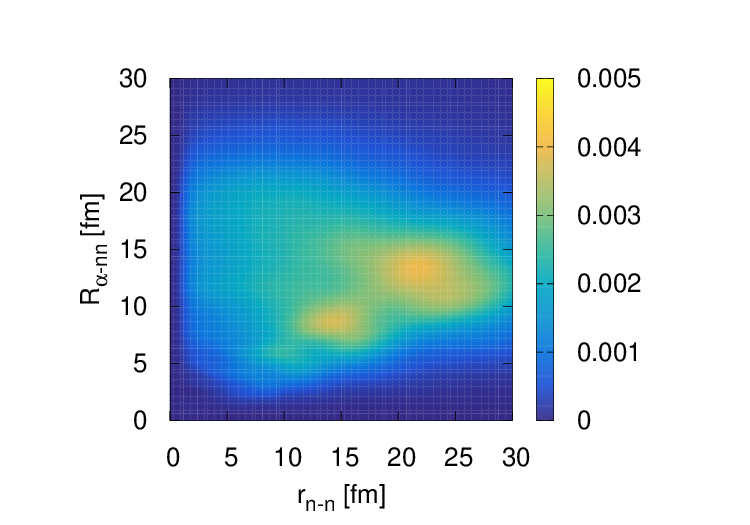}
\caption{
Time evolution of the density distribution $D(r_{n\text{-}n},R_{\alpha\text{-}nn},t)$ in units of $e^2$ for the E1-excited state of $^{6}\mathrm{He}$ in the Jacobi coordinate system describing the relative motion of the two neutrons. The density distributions are shown at $ct = 50$, $150$, and $300~\mathrm{fm}$ from left to right.
}
\label{fig:r3R3_evolution}
\end{figure*}

We next investigate the time evolution of the density distribution in the Jacobi coordinate set ($r_{n\text{-}n}, R_{\alpha\text{-}nn}$) describing the relative motion of the two neutrons. As shown in Fig.~\ref{fig:r3R3_evolution}, the density distribution at $ct = 50~\mathrm{fm}$ (left) still exhibits a concentration at small $r_{n\text{-}n}$, indicating that the dineutron-like configuration persists at early times. As time evolves to $ct = 150~\mathrm{fm}$ (center), the density extends toward larger values of $r_{n\text{-}n}$, showing that the neutron--neutron correlation is gradually weakened. At $ct = 300~\mathrm{fm}$ (right), the density is widely spread in $r_{n\text{-}n}$, indicating that the two neutrons propagate apart in the continuum. The distribution also extends in $R_{\alpha\text{-}nn}$, reflecting the overall expansion of the system with low density.

\begin{figure*}[t]
\centering
\includegraphics[trim=30 0 30 0, width=0.32\textwidth]{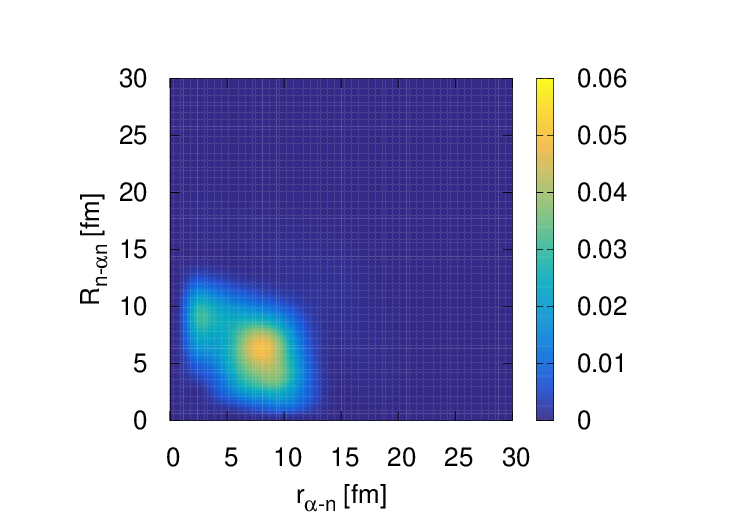}
\includegraphics[trim=30 0 30 0, width=0.32\textwidth]{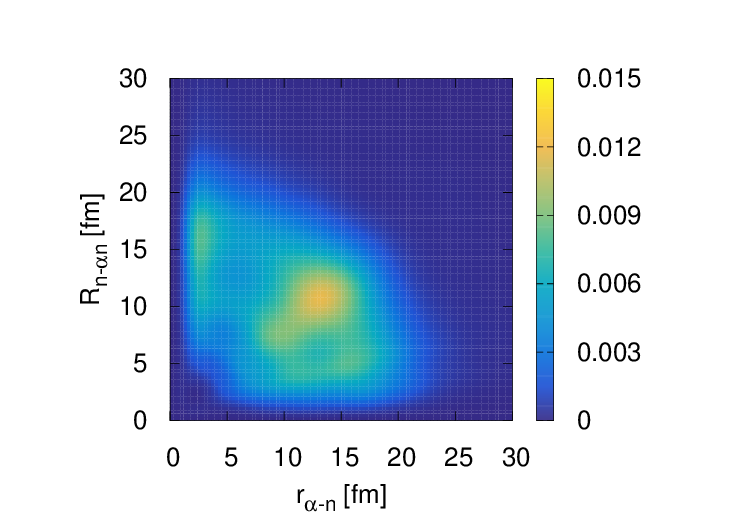}
\includegraphics[trim=30 0 30 0, width=0.32\textwidth]{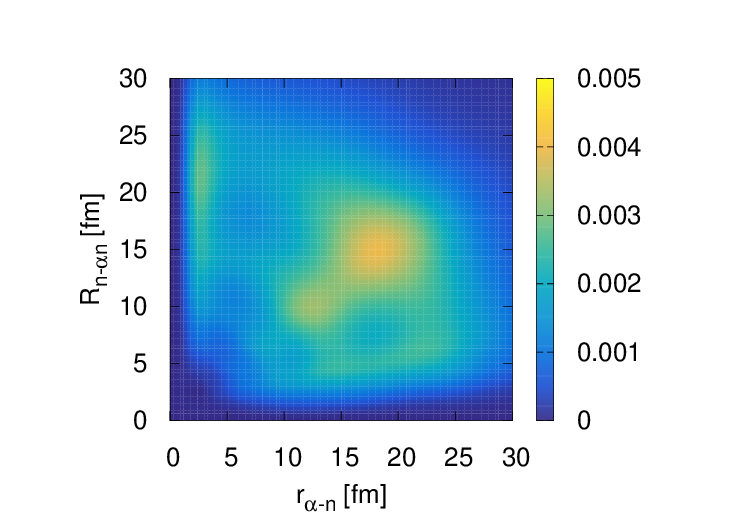}
\caption{
Same as Fig.~\ref{fig:r3R3_evolution}, but for the density distribution $D(r_{\alpha\text{-}n}, R_{n\text{-}\alpha n}, t)$ in units of $e^2$ in the Jacobi coordinate system corresponding to the core--neutron relative motion.
}
\label{fig:r1R1_evolution}
\end{figure*}

Finally, we examine the time evolution in another Jacobi coordinate set $(r_{\alpha\text{-}n},R_{n\text{-}\alpha n})$ corresponding to the core--neutron relative motion, as shown in Fig.~\ref{fig:r1R1_evolution}.
At $ct = 50$~fm, the density distribution exhibits a localized structure.
In addition, a component is visible in the region of small $r_{\alpha\text{-}n}$.
As time evolves, the density extends toward larger values of both $r_{\alpha\text{-}n}$ and $R_{n\text{-}\alpha n}$, indicating the development of extended breakup configurations.
At $ct = 300$~fm, the density becomes broadly distributed over both coordinates, showing that all constituents move apart in the continuum.

Two characteristic structures can be identified in the density distribution in the Jacobi coordinate set $(r_{\alpha\text{-}n}, R_{n\text{-}\alpha n})$.
The first structure corresponds to sequential decay, and it appears as a component localized at small $r_{\alpha\text{-}n}$ while extending toward larger $R_{n\text{-}\alpha n}$, showing a vertical line-like structure in the figure and indicating the formation of a core--neutron subsystem as a $^5$He resonance accompanied by the outward propagation of the remaining neutron.
A similar feature is also observed at small $R_{n\text{-}\alpha n}$, which can be attributed to antisymmetrization between the two neutrons.
In contrast, the second structure corresponds to direct breakup and is characterized by a simultaneous extension in both $r_{\alpha\text{-}n}$ and $R_{n\text{-}\alpha n}$, indicating that all constituents separate without forming an intermediate subsystem.
Notably, the sequential decay component is already visible at an early stage of the time evolution ($ct=50$~fm), indicating the early formation of a core--neutron subsystem.

These results clearly indicate that multiple decay modes coexist in the E1 excitation of $^6$He during the time evolution.
In particular, the analysis of the time evolution with the present formalism enables a clear identification and visualization of the sequential decay process, while also describing the transition toward more spatially extended continuum configurations.
This demonstrates that the present approach provides a unified description of decay dynamics in the three-body continuum.

Overall, the present results provide a clear picture of the time evolution of the E1-excited three-body system.
The initial state is characterized by a dineutron-like configuration, which evolves into spatially extended continuum states.
The analysis in different Jacobi coordinate systems reveals the coexistence of multiple decay modes, in particular sequential decay and direct breakup.
The sequential decay component can be clearly identified, and its emergence is already observed at an early stage of the time evolution.
These results demonstrate not only that the present formalism can visualize the emergence of different decay modes in time, but also that such dynamical information can be extracted within the same CSM framework that has previously been used for resonance spectroscopy, continuum responses, and scattering observables.

\section{Summary}

In this work, we have developed a time-evolution formalism in the complex scaling method based on the extended completeness relation, in which the central quantity is the complex-scaled time-evolution operator. The present formulation extends the scope of the CSM, which has been used for resonances, continuum responses, and scattering observables, to the explicit description of time evolution in the continuum.

The validity of the method has been confirmed through a test calculation for a simple model system. The comparison with the Crank--Nicolson solution of the time-dependent Schr\"odinger equation shows that the present approach accurately reproduces the time evolution of the wave packet, including the contributions from broad resonance states.

We have applied the present formulation to the E1 excitation of $^6$He within a three-body model. The initial E1-excited state exhibits a dineutron-like configuration, which gradually evolves into spatially extended continuum states. The analysis of the density distributions in different Jacobi coordinate systems reveals the coexistence of multiple decay modes, including sequential decay and direct breakup, together with a global expansion of the wave packet.

These results demonstrate that the present approach provides a direct and unified description of decay dynamics in weakly bound systems. In particular, the present formalism enables a clear visualization of how different decay modes emerge and evolve in time, establishing a direct link between initial-state correlations and asymptotic continuum dynamics.

The present work thus broadens the scope of the CSM from a framework formulated mainly in the energy representation to one that also describes continuum dynamics in the time domain. In this sense, the present formalism provides a missing dynamical component of the CSM and offers a unified perspective on structure, continuum, and decay in weakly bound and unbound nuclear systems.

\begin{acknowledgments}
One of the authors (Y.K.) would like to thank Prof.~T.~Uesaka and Dr.~K.~Yoshida for valuable discussions.
The work shown in this paper was supported by JSPS KAKENHI Grants No. JP22K03643, No. JP25H01268, and JST ERATO Grant No. JPMJER2304, Japan.
\end{acknowledgments}

\bibliographystyle{apsrev4-2}

\bibliography{references}

\end{document}